\documentclass[twocolumn,superscriptaddress,floatfix,preprintnumbers,aps,prl,10pt]{revtex4-2}

\usepackage[colorlinks=true,breaklinks=true]{hyperref}
\usepackage[utf8]{inputenc}
\usepackage{longtable}

\hypersetup{urlcolor=Blue,
	    citecolor=Blue,
	    linkcolor=Blue}
\usepackage{url}
\usepackage{mathtools}
\usepackage{amsmath}
\usepackage[dvipsnames]{xcolor}

\usepackage{orcidlink}
\usepackage{JournalAbbreviations}
\usepackage{microtype}

\newcommand{\unitspace}{\ensuremath{\mskip\thinmuskip}}

\newcommand{\gccm}{\ensuremath{{\mathrm{g}\unitspace\mathrm{cm}^{-3}}}}

\newcommand{\nux}{\ensuremath{\nu_{x}}}

\begin{document}

\title{Gravitational-Wave Signatures of Nonstandard Neutrino Properties\\ in Collapsing Stellar Cores}

\author{Jakob Ehring~\orcidlink{0000-0003-2912-9978}}
\affiliation{Institute of Physics, Academia Sinica, No.~128, Section~2, Academia Road, 115201 Taipei City, Taiwan}
\affiliation{Max-Planck-Institut f\"ur Physik, Boltzmannstraße~8, 85748 Garching, Germany}
\affiliation{Max-Planck-Institut f\"ur Astrophysik, Karl-Schwarzschild-Straße~1, 85748 Garching, Germany}

\author{Sajad Abbar~\orcidlink{0000-0001-8276-997X}}
\affiliation{Max-Planck-Institut f\"ur Physik, Boltzmannstraße~8, 85748 Garching, Germany}

\author{Hans-Thomas Janka~\orcidlink{0000-0002-0831-3330}}
\affiliation{Max-Planck-Institut f\"ur Astrophysik, Karl-Schwarzschild-Straße~1, 85748 Garching, Germany}

\author{Georg Raffelt~\orcidlink{0000-0002-0199-9560}}
\affiliation{Max-Planck-Institut f\"ur Physik, Boltzmannstraße~8, 85748 Garching, Germany}

\author{Ko Nakamura~\orcidlink{0000-0002-8734-2147}}
\affiliation{Department of Applied Physics, Fukuoka University, Nanakuma Jonan 8-19-1, Fukuoka 814-0180, Japan}

\author{Kei Kotake~\orcidlink{0000-0003-2456-6183}}
\affiliation{Department of Applied Physics, Fukuoka University, Nanakuma Jonan 8-19-1, Fukuoka 814-0180, Japan}
\affiliation{Institute for Theoretical Physics, University of Wroclaw, 50-204 Wroclaw, Poland}

\begin{abstract}
We present a novel multimessenger approach for probing nonstandard neutrino properties through the detection of gravitational waves (GWs) from collapsing stellar cores and associated supernova explosions. We show that neutrino flavor conversion inside the proto-neutron star (PNS), motivated by physics Beyond the Standard Model (BSM), can significantly boost PNS convection. This effect leads to large-amplitude GW emission over a wide frequency range during an otherwise relatively quiescent GW phase shortly after core bounce. Such a signal provides a promising new avenue for exploring nonstandard neutrino phenomena and other BSM physics impacting PNS convection.
\end{abstract}

\maketitle

\textit{\textbf{Introduction.}}---Since their first detection on September 14, 2015, by the LIGO and Virgo collaborations~\cite{LIGOScientific:2016aoc}, GWs from compact objects have become crucial tools for probing new physics under extreme conditions~\cite{Bauswein+2010, Yunes:2013dva, Orsaria+2019, Tsang+2019, Andersson2019, Weih+2020, Bauswein+2020}.
In this \textit{Letter}, we demonstrate that GWs produced by collapsing stellar cores and associated supernova (SN) explosions can serve as powerful probes of new physics that influences convection within the hot PNS, with a particular focus on nonstandard neutrino properties.

In the cores of collapsing stars, asymmetric motions of large masses with high velocities leave an imprint on spacetime in the form of GWs.
While such signals from SNe have yet to be observed, neutrino-hydrodynamic simulations predict several contributions.
In nonrotating progenitors, an initial GW burst arises from ``prompt'' convection behind the decelerating SN shock shortly \hbox{after} bounce, caused by a negative entropy gradient.
This activity calms down after a few 10\,ms, followed by a generic phase (around 100\,ms) of fairly low GW emission.

Subsequently, violent mass motions connected to hydrodynamic instabilities in the neutrino-heated postshock layer and PNS oscillations, stimulated either by mass downflows from outside or by convection inside the PNS, become the primary sources.
Later, GWs also come from long-lasting PNS convection, accretion of fallback mass, and the asymmetric morphology of expanding explosion ejecta and emitted neutrinos, both leading to a long-term GW strain or ``memory'' \cite{Braginsky+1987, Christodoulou1991, Favata2010}.
In rotating, collapsing stellar cores,  additional sources include the core deformation at bounce, causing different characteristic burst signals, triaxial hydrodynamic (spiral and bar-mode) and magneto-hydrodynamic instabilities, and mass ejection in polar jets (see \cite{Ott2009, Kotake2013, KotakeKuroda2017_Handbook, Kalogera+2021, Abdikamalov+2022, Arimoto+2023, Mezzacappa+2024} for reviews).

Neutrino flavor conversion (FC) in the SN environment introduces a major uncertainty in current SN modeling.
The high neutrino densities lead to complex, nonlinear, collective FC scenarios that can spawn a range of intriguing phenomena \cite{Pastor:2002we, Duan:2006an, Duan:2010bg, Chakraborty:2016yeg, Volpe:2023met}.
Despite growing evidence of their relevance for SN dynamics, FC has been ignored in most simulations.
It is only recently that multidimensional neutrino-hydrodynamic SN simulations have started to include schematic implementations \cite{Ehring2023a, Ehring2023b, Nagakura:2022kic,Nagakura:2023PhRvL.130u1401N}.
Specifically, the phenomenon of \emph{fast} FC (FFC) was considered, where FC might occur on much smaller scales than other phenomena in the SN core.
These studies indicate that FCs can both support or suppress neutrino-driven explosions, depending on the progenitor and the FC region~\cite{Ehring2023b}.

However, FFCs thus explored \cite{Ehring2023a, Ehring2023b} are not expected {\em inside\/} the PNS, at least not during the first few 100\,ms after bounce~\cite{Glas2020a, Abbar:2019zoq, Nagakura:2021hyb}.
On the other hand, BSM neutrino properties provide several mechanisms for generic FC within a PNS.
For instance, resonant conversion and reconversion between active and sterile neutrinos can accelerate the PNS's neutrino loss \cite{Hidaka:2007se, Hidaka:2006sg, Caldwell:1999zk}.
Nonstandard neutrino self-interactions ($\nu$NSSI) \cite{Bialynicka-Birula:1964ddi, Bardin:1970wq, Berryman:2022hds} can lead to neutrino flavor equipartition inside the PNS~\cite{Abbar:2022jdm}.
Likewise, BSM magnetic transition moments for Majorana neutrinos can also cause some sort of equipartition between neutrinos and antineutrinos, if there are strong magnetic fields in the PNS~\cite{Abbar:2020ggq, Kharlanov:2020cti, Sasaki:2021bvu}.
Such an equipartition could significantly impact the thermodynamics of SNe~\cite{Suliga:2024oby}.

\begin{figure}[t]
    \includegraphics[width=1.0\columnwidth]{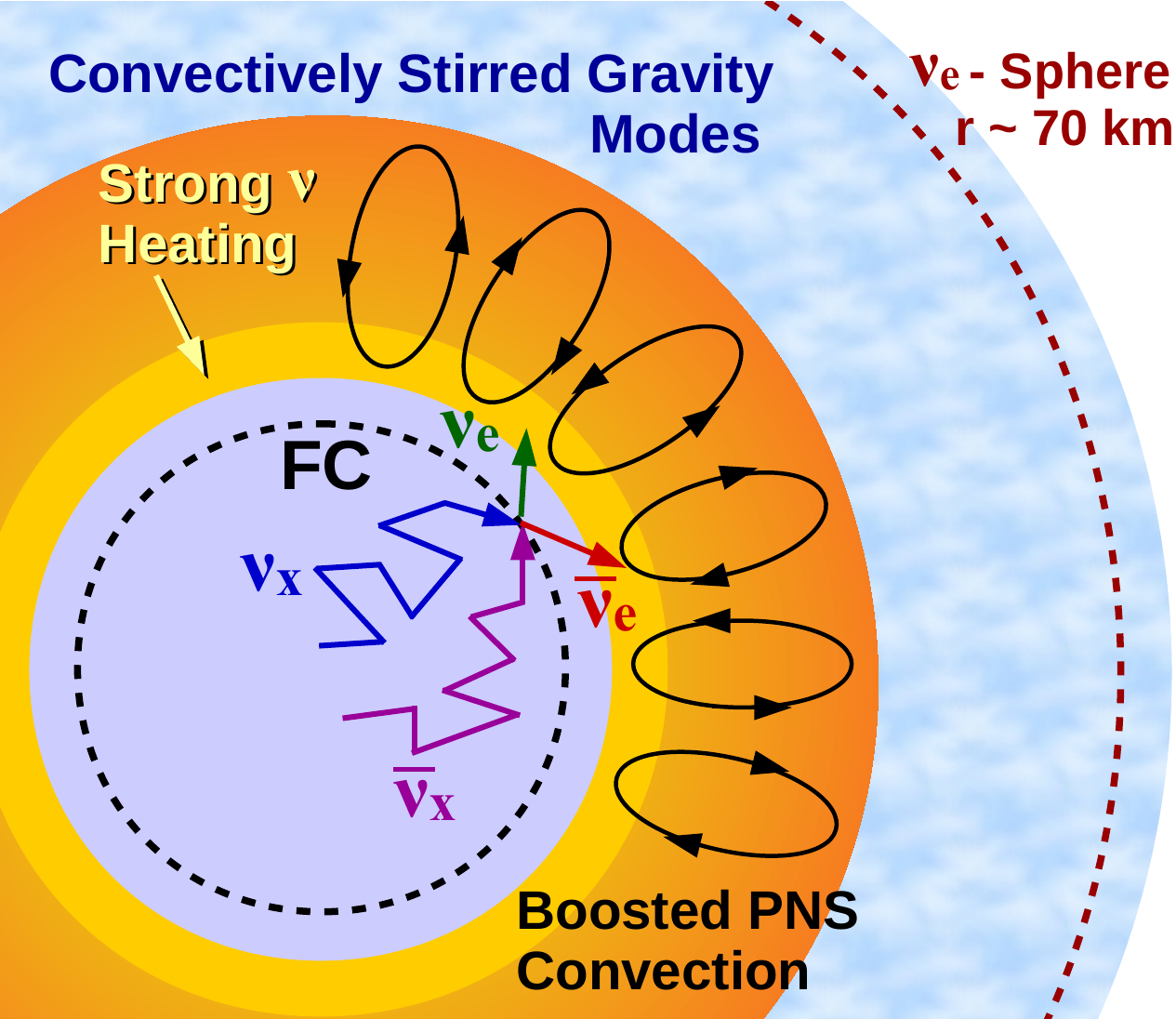}
    \caption{
    FC of high-energy $\nu_x$ and $\bar\nu_x$ create $\nu_e$ and $\bar\nu_e$ in the PNS interior, an effect assumed to occur at the black dashed circle.
    If this happens well inside the average electron-type neutrinosphere, the newly created $\nu_e$ and $\bar\nu_e$ are quickly absorbed by free nucleons.
    This strong local heating (bright orange layer) enhances PNS convection (orange layer).
    The convective shell as well as g-mode activity (so-called gravity waves, not to be confused with gravitational waves) instigated in the convectively stable near-surface layer emit GWs.
    }
    \label{fig:schematic}
\end{figure}

Here we consider FFC as a prototypical example and show that FC inside the PNS, at densities above a few $10^{12}$\,g\,cm$^{-3}$, visibly intensifies the GW signal shortly after bounce, during the otherwise relatively GW-quiescent phase.
This enhancement arises from strong, localized heating caused by the conversions of $\nu_x$ and $\bar\nu_x$  $(x = \mu, \tau)$  to $ \nu_e$ and $ \bar\nu_e$, which are immediately absorbed in the medium.
This increased heating leads to strong convective activity, which in turn can excite hydrodynamic gravity modes (g-modes) near the PNS surface, resulting in strong GW emission.
This effect is schematically illustrated in Fig.~\ref{fig:schematic}.

\textit{\textbf{Simulation Setup and GW Analysis.}}---The considered simulations were discussed in Ref.~\cite{Ehring2023b}.
They were conducted with the neutrino-hydrodynamics code \textsc{Alcar} \cite{Just2015a,Just2018a}, assuming axial symmetry (2D).
\textsc{Alcar} is a state-of-the-art, Eulerian, conservative, higher-order Godunov-type finite-volume solver designed for both 1D and multi-D nonrelativistic fluid dynamics, coupled with a two-moment scheme to treat energy-dependent, three-flavor neutrino transport.
General relativistic corrections for gravity and transport are included.
The key features of the code and the implementation details of the neutrino FCs are described in Ref.~\cite{Ehring2023a}.

The FC implementation in Refs.~\cite{Ehring2023a, Ehring2023b} was motivated by FFCs, but the parametric representation can be interpreted in terms of any BSM physics that would cause similar effects.
For examples, see Supplemental Material (SM)\cite{SupplementalMaterial}, which also includes Refs.~\cite{Blennow:2008er, Chang:2022aas, Fiorillo2024a, Fiorillo2024c, Bhattacharya:2023wzl, Warren:2014qza, Warren:2016slz, Syvolap:2019dat, Suliga:2020vpz, Ray:2023gtu, Ray:2024jeu, Courant1928a, Kuroda+2014, Murphy+2009, Shibagaki2021a, Andresen2017a, Acernese2015a, LSC2015a, Abbott2020c, Aso2013a, Michimura2020a, Hild2008_1, Srivastava2022a, Flanagan1998a}.
The following recipe was used:
(1)~we assume that pairwise $\nu\bar\nu$ FCs are ``instantaneous,'' meaning that they take place over spatial scales much smaller than the numerical grid cells and over time scales much shorter than the computational time steps~\cite{Xiong:2024pue}.
This description is also reasonable for BSM-induced FCs, where the scales are governed, for example, by the BSM neutrino interaction potential or the magnetic field strength~\cite{Abbar:2022jdm, Abbar:2020ggq, Kharlanov:2020cti, Sasaki:2021bvu}.
(2)~We assume that the FCs result in flavor equilibrium under the constraint that lepton number is individually conserved for each flavor, particularly for electron-type neutrinos.
BSM physics could violate this restriction, potentially allowing for full flavor equilibration, meaning that our more constraining assumption is conservative.
(3)~Our FC treatment conserves total energy and momentum, while respecting the Pauli exclusion principle.
(4)~We activate FCs throughout the region where the matter density $\rho$ is below a chosen threshold value, i.e., $\rho < \rho_\mathrm{c}$.
Although Ref.~\cite{Xiong:2024pue} backs up a scale-separated treatment for FFC even inside the neutrinosphere, such an effective transport has its limitations~\cite{Urquilla+2025}.
Especially, classical transport cannot capture equilibria with non-zero coherence as pointed out in~\cite{Froustey+2025} for collisional $\nu$FC.
Our approach should be understood as explorative and generic for FCs occurring on much smaller scales as motivated by $\nu$NSSI and other BSM scenarios with similar effects (for details and resolution tests, see SM~\cite{SupplementalMaterial}).

The simulations were initially evolved in 1D until core bounce and then mapped onto a 2D polar coordinate grid with 640 logarithmically spaced radial zones and 90 equally spaced angular zones.
The central core of 2\,km was kept 1D, thus permitting larger time steps without significant effects on the evolution.
To initiate nonradial hydrodynamic instabilities, which otherwise arise only from uncontrolled numerical noise, we perturbed the local density in every cell of the computational grid during the mapping with a random amplitude of up to 0.1\,\%.

We selected three progenitors with different zero-age main-sequence masses: exploding 9\,M$_\odot$~\cite{Woosley2015a, Radice2017a, Burrows2020a, Burrows2021a, Just2018a, Glas2019a, Stockinger2020a} and 11.2\,M$_\odot$ models~\cite{Woosley2002a, Buras2006b, Marek2009b, Takiwaki2012a, Mueller2015b}, and a nonexploding 20\,M$_\odot$ model~\cite{Woosley2007a, Melson2015a, Just2018a, Vartanyan2018a, Glas2019a}.
Our corresponding core-collapse simulations are therefore named M9, M11.2, and M20.
For cases without FC, the name is supplemented with ``noFC'' or else with the choice for the FC threshold density $\rho_\mathrm{c}$.
The 2D models considered here differ slightly from those of \cite{Ehring2023b} in being mapped to 2D 5\,ms earlier, which is relevant for prompt postshock convection, and including the many-body corrections discussed in \cite{Horowitz2017a}.

GW signals are postprocessed, employing a variant of the standard quadrupole formula [e.g., Eq.~(34) in Ref.~\cite{Finn1990a}; see also SM~\cite{SupplementalMaterial}].
The algorithm makes use of the Euler equations to avoid the second time derivative and associated numerical errors.
Spectrograms are calculated from the strain using a short-time Fourier transformation with a sliding window of 20\,ms.
Due to axisymmetry in 2D models, the cross mode is zero in all directions and both cross and plus modes vanish for polar observers.
Therefore we only report on the plus mode, $h_{+}$, observed in the equatorial plane at a distance of 10\,kpc.

\begin{figure}
     \includegraphics[width=1.0\columnwidth]{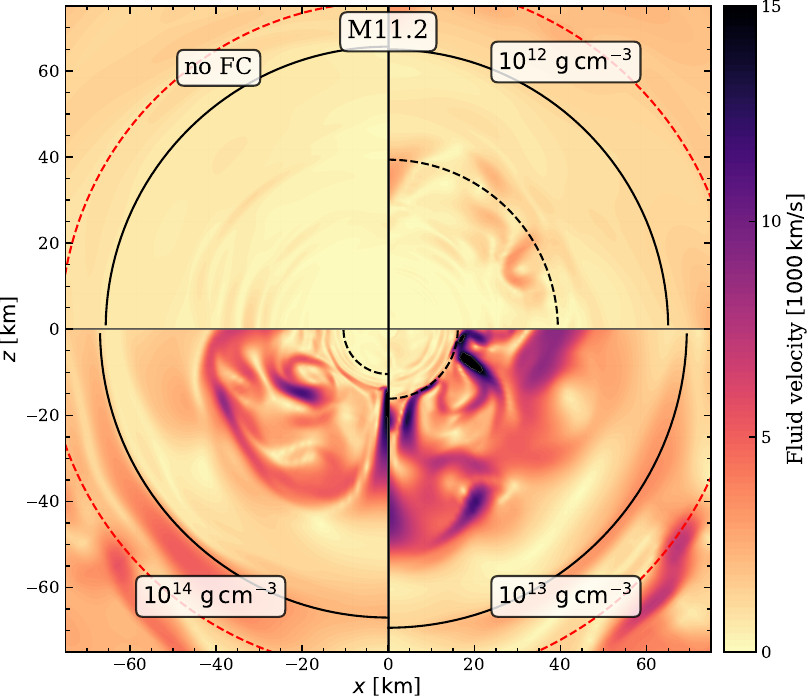}
     \vskip-4pt
    \caption{
    Convective mass motions inside the PNS for our 2D SN model M11.2 of an 11.2\,M$_\odot$ progenitor with different neutrino FC scenarios.
    The four quadrants display the color-coded magnitude of the stellar gas velocity 40\,ms after core bounce for the model without FCs (labeled by noFC) and values of $\rho_c=10^{12}$, $10^{13}$, and $10^{14}$\,g\,cm$^{-3}$ for the threshold density below which FC is assumed to occur (clockwise from top left).
    The black dashed circular lines indicate these inner boundaries of the FC regions.
    Convection is almost nonexistent within the PNS at this early time after bounce in model noFC, whereas in all cases with FCs, significantly stronger convective activity has already developed.
    The red dashed circular lines indicate the average energy spheres of $\nu_e$, and the solid black circular lines mark the locations where the matter density is $10^{11}$\,g\,cm$^{-3}$, which effectively coincide with the average energy spheres of $\bar\nu_e$.
    }
    \label{fig:conv}
    \vskip-8pt
\end{figure}

\textit{\textbf{Results and Discussion.}}---A dynamically relevant consequence of FC inside the PNS is enhanced convective activity.
This triggers and strengthens GW emission.
The convective velocities become several times higher than without FC.
At densities exceeding a few $10^{12}$\,g\,cm$^{-3}$, weak processes hinder the presence of $\bar\nu_e$ due to $e$ and $\nu_e$ degeneracy.
Therefore, the dominant FC channel at high $\nu_x$ energies is $\nu_x, \bar\nu_x \rightarrow \nu_e, \bar\nu_e$.
The newly produced $\nu_e$ and $\bar\nu_e$ are quickly absorbed by the surrounding medium (see \cite{Ehring2023a} for a detailed discussion of this effect in 1D).
Additional, local heating fosters vigorous convective mass motions inside the PNS, as illustrated by Fig.~\ref{fig:schematic} and visible in our simulations (Fig.~\ref{fig:conv}).

\begin{figure*}
    \includegraphics[width=1.\textwidth]{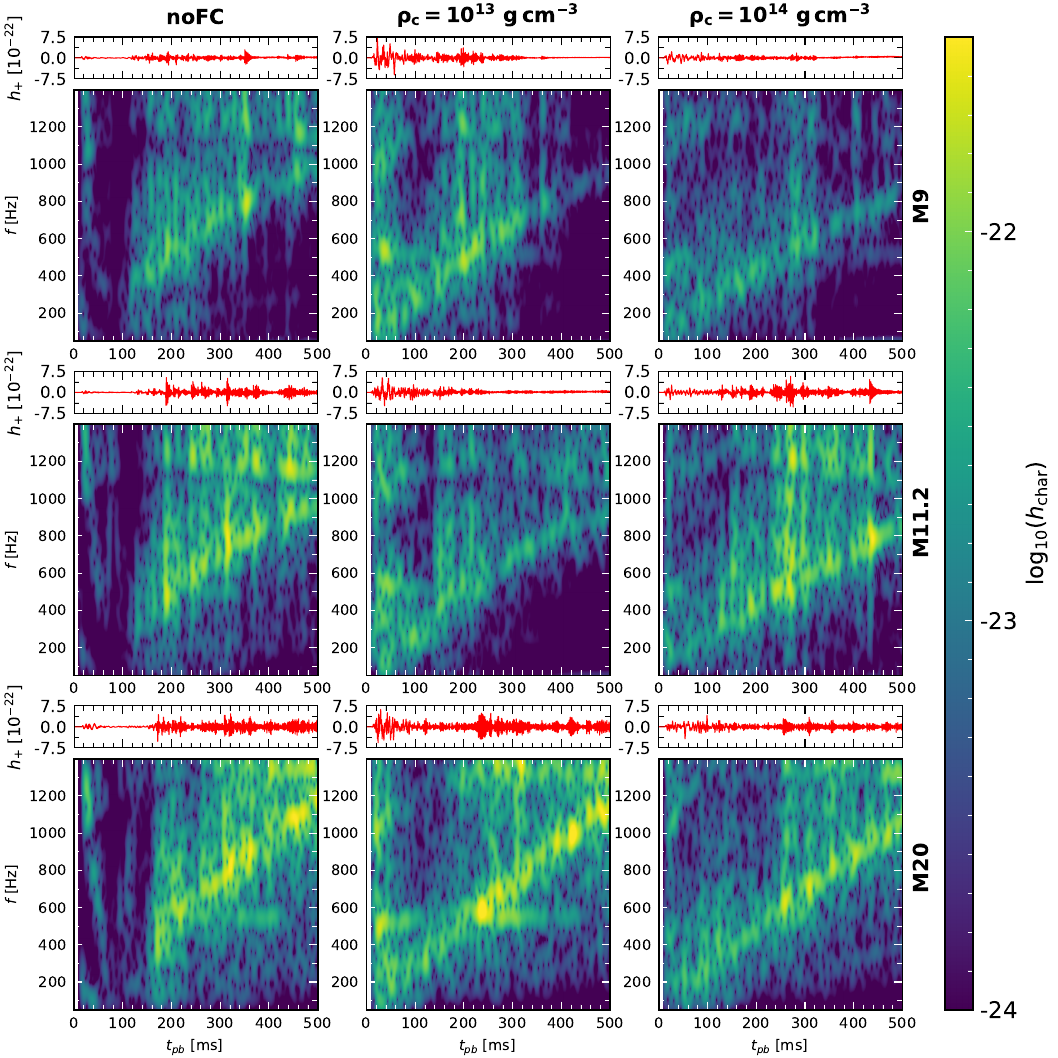}
    \caption{
    GW strains and their spectrograms for a SN distance of 10\,kpc vs.\ time after bounce and an observer at the equator.
    The different panels correspond to our 2D SN simulations of 9\,M$_\odot$ ({\em top row}), 11.2\,M$_\odot$ ({\em middle row}), and 20\,M$_\odot$ progenitors ({\em bottom row}), considering different FC scenarios: noFC ({\em left panels}), FC at $\rho < \rho_\mathrm{c}= 10^{13}$\,g\,cm$^{-3}$ ({\em middle column}), and at $\rho < \rho_\mathrm{c}= 10^{14}$\,g\,cm$^{-3}$ ({\em right panels}).
    Our noFC models possess extended periods (lasting 70--100\,ms) of relative quiescence after a short post-bounce phase of GW activity due to prompt postshock convection, whereas the models with FCs exhibit strong GW emission over a wide frequency range during this time interval.
    Analogous results for $\rho_\mathrm{c}= 10^{9}, 10^{10}, 10^{11}$, and $10^{12}$\,g\,cm$^{-3}$ are provided in SM~\cite{SupplementalMaterial}.
    }
    \label{fig:GW}
\end{figure*}

The intense convection caused by FCs in the PNS excites gravity-mode (g-mode) oscillation of the convectively stable outer PNS layer.
Both PNS convection and g-mode activity generate large-amplitude GW emission over a broad frequency range, occurring shortly after core bounce, in particular for our models with $\rho_\mathrm{c} = 10^{13}$ and $10^{14}\,\gccm$, as visible in Fig.~\ref{fig:GW}.

Collapsing nonrotating stars without neutrino FCs in the PNS interior exhibit a short post-bounce pulse of GW emission caused by an initial phase of prompt postshock convection due to negative entropy and electron-fraction gradients behind the decelerating core-bounce shock.
This phase of 20--50\,ms is followed by a more quiescent period of 50--100\,ms with significantly weaker GW activity, before the GW emission amplifies again due to the onset of violent mass motions in the postshock layer, connected to large-scale shock-deformation modes and convective overturn triggered by neutrino heating behind the stalled SN shock (left panels of Fig.~\ref{fig:GW} and \hbox{Refs.~\cite{Marek+2009,Murphy+2009,Yakunin+2010,Mueller+2013,Yakunin+2015,Morozova+2018}).}

The quiescent period is absent in our models with FCs below $\rho_\mathrm{c} = 10^{13}$ and $10^{14}\,\gccm$.
Instead, the GW emission setting in right after bounce has a more than 10 times larger GW strain $h_+$ than the noFC case (see Fig.~\ref{fig:GW} for the strain at source distance $D = 10$\,kpc) with quadrupole amplitudes $A_{20}^\mathrm{E2} = 8\sqrt{\pi/15}Dh_+$ up to 50--110\,cm, depending on the model, for observers at 90$^\circ$ inclination angle to the 2D symmetry axis.
This is the strongest GW signal witnessed for nonrotating stellar collapse in 2D so far, and it bridges the otherwise calmer phase with a continuous high level of GW activity.

This finding is generally valid without stellar rotation, although the GW amplitudes from prompt convection, the subsequent, more silent phase, and the pre-explosion postshock instabilities vary substantially between different progenitors, explosion behaviors, and modeling details \cite{Mueller+2004,Marek+2009,Murphy+2009,Yakunin+2010,Mueller+2013,Yakunin+2015,Morozova+2018,Jardine+2022,Murphy+2024}.
Our GW signal from FC-induced PNS convection is somewhat weaker for $\rho_\mathrm{c} = 10^{14}$ than $\rho_\mathrm{c} = 10^{13}\,\gccm$, because the $\nux$ fluxes are smaller deeper inside and $\nu_x, \bar\nu_x \rightarrow \nu_e, \bar\nu_e$ FC at higher densities decelerates neutrino diffusion out of the PNS core.
Models with $\rho_\mathrm{c} = 10^{12}\,\mathrm{g\,cm^{-3}}$ show only moderately increased PNS convection (Fig.~\ref{fig:conv}) and GW amplitudes (see SM~\cite{SupplementalMaterial}), since the lower $\nu_e$ degeneracy favors the $\nu_e, \bar\nu_e \rightarrow \nu_x, \bar\nu_x$ channel over $\nu_x, \bar\nu_x \rightarrow \nu_e, \bar\nu_e$.

Although the basic GW components in nonrotating 2D and 3D models are similar, their amplitudes are 10--20 times lower in 3D \citep{Andresen2017a, Radice+2019, Powell+2019, Powell+2020, Mezzacappa2020a, Mezzacappa+2023, Vartanyan+2023}.
Assessing their detectability therefore requires 3D models.
Thus downscaling the initial FC-induced 50--100\,ms GW pulse of our models in Fig.~\ref{fig:GW}, we estimate that it might be captured by current LIGO-Virgo-KAGRA searches for galactic SNe up to roughly a kpc with signal-to-noise ratios of around 10.
In view of their relatively broad frequency range, these GWs may become observable to $D > 10$\,kpc by upcoming interferometers such as the Einstein Telescope and the Cosmic Explorer~\cite{Punturo:2010zz, Reitze:2019iox}, potentially including the entire Milky Way (for details, see SM~\cite{SupplementalMaterial}).

To link the particle-physics related burst to the phase preceding the otherwise quiescent window, the core-bounce time needs to be accurately determined by measuring the neutronization burst in current SN neutrino detectors; and to discriminate the pulse from the bounce and post-bounce signal of rapidly rotating stellar cores \cite{Dimmelmeier+2008,Mueller+2004,Abdikamalov+2014,Powell+2020}, other characteristic GW signatures associated with high angular momentum must be identified, e.g., frequency shifts of the dominant characteristic quadrupolar g- and f-mode PNS oscillations \cite{Powell+2020,Jardine+2022,Powell+2024} or narrow-band, quasiperiodic waveforms due to spiral modes~\cite{Kuroda+2014,Shibagaki+2020} that are possibly correlated with rotation-induced modulations of the neutrino signal \cite{Takiwaki+2018,Shibagaki2021a,Takiwaki+2021}.

Thus GWs may be used in a multimessenger approach as one of the most sensitive probes of nonstandard neutrino properties under extreme conditions and, indirectly, even of neutrino self-interactions.
Otherwise, if an FC-induced GW burst is not observed, very stringent constraints on nonstandard neutrino properties could be placed.
For instance, resonant conversions of sterile neutrinos inside the PNS~\cite{Hidaka:2007se, Hidaka:2006sg, Caldwell:1999zk} would be disfavored for parameters where enhanced PNS convection is triggered.
More such possibilities are discussed in SM~\cite{SupplementalMaterial}.

Our FC treatment assumed lepton number conservation for each flavor individually.
However, this assumption can be maximally violated in certain BSM models, potentially achieving total flavor equipartition~\cite{Abbar:2022jdm, Abbar:2020ggq, Kharlanov:2020cti, Sasaki:2021bvu}.
Such scenarios with stronger FC than in our conservative prescription might amplify the effects of PNS convection, because even without maximal FC the impact is significant according to our results.
This highlights the broader relevance of our findings.
However, due to the nonlinear and coupled nature of SN physics, extrapolations are insufficient, but detailed simulations are necessary for each specific FC scenario to fully assess its consequences.

While neutrino FC scenarios can be studied using the neutrino signal from a future galactic SN, our proposed GW analysis offers an intriguing tool to probe the dynamical effects of FC.
In fact, the neutrino signal can only reveal the final state of neutrinos as they exit the SN, making it almost impossible to pinpoint the location and physics causing FC.
For example, FC due to $\nu$NSSI or sterile neutrinos can also take place farther away from the PNS, but with the GW analysis one can gain insights whether the FC occurred within the PNS.

\textit{\textbf{Summary and Outlook.}}---We have investigated the GW signals generated by 2D simulations of stellar collapse, specifically focusing on the presence of neutrino FC inside PNSs.
Our findings indicate that such FCs can lead to a new component in the GW signal during an otherwise much more quiescent phase of GW emission.
This phenomenon arises from enhanced local heating in the PNS interior due to absorption of $\nu_e$ and $\bar\nu_e$ produced via the $\nu_x, \bar\nu_x \rightarrow \nu_e, \bar\nu_e$ FC channel.
This boosts PNS convection and thus excites g-modes in the near-surface layer of the PNS, causing the strongest 50--100\,ms long GW pulse witnessed in nonrotating SN models shortly after core bounce so far, followed by GW activity that bridges the otherwise GW-silent phase between prompt postshock convection and hydrodynamic instabilities in the neutrino-heated postshock layer.

Given its broad frequency range, the FC-induced GW signal with amplitudes expected for corresponding 3D sources may be detectable from distances up to a kpc with current facilities and out to $D > 10$\,kpc with future laser interferometers.
If such a signal is not observed, stringent constraints on some BSM neutrino scenarios including nonstandard neutrino interactions could be inferred.
This strategy offers one of the most promising avenues for exploring nonstandard neutrino physics under extreme conditions with neutrino and GW information.

While in this study we considered neutrino FC enhancing convection inside the PNS, its convection may also be affected by other BSM physics.
Of particular interest is, e.g., axion emission~\cite{Fiorillo2023f, Caputo+2024}.
Though such an energy-loss channel is, in principle, expected to accelerate PNS neutrino cooling, nonlinear feedback between the additional particle cooling and convection could yield unexpected effects, also modifying the GW emission.
We leave such possibilities for future dedicated investigations.

Although our focus was on GWs from collapsing stars, BSM physics could also impact GW emission from binary NS mergers (NSMs).
Specifically, BSM particle production might influence the characteristic ring-down peak feature in NSM GW spectra \cite{Oechslin+2007, Stergioulas+2011, Bauswein+2012, Takami+2014}, shifting the peak frequency or damping the quadrupole oscillations of the NSM remnant that are responsible for this feature.

In view of our promising results, a wide variety of questions demand further exploration.
The incorporation of nonstandard physics in 3D SN simulations is crucial to predict the exact properties and detectability distance of the signals, because the GW strain is known to be systematically lower in 3D compared to 2D \citep{Andresen2017a, Radice+2019, Powell+2019, Powell+2020, Mezzacappa2020a, Mezzacappa+2023, Vartanyan+2023}.

\textit{\textbf{Acknowledgments.}}---This work was supported by the German Research Foundation (DFG) through the Collaborative Research Centre ``Neutrinos and Dark Matter in Astro- and Particle Physics (NDM),'' Grant No.\ SFB-1258\,--\,283604770, and under Germany’s Excellence Strategy through the Cluster of Excellence ORIGINS EXC-2094-390783311.
J.E.\ received funding as International Research Fellow of the Japan Society for the Promotion of Science (Postdoctoral Fellowships for Research in Japan).
K.N.\ and K.K.\ were supported in part by Grants-in-Aid for Scientific Research of the Japan Society for the Promotion of Science (JSPS, Nos.\ JP23K20862, JP23K22494,  and JP24K00631) and the funding from Fukuoka University (Grant GR2302), and JICFuS as ``Program for Promoting Researches on the Supercomputer Fugaku'' (Structure and Evolution of the Universe Unraveled by Fusion of Simulation and AI; Grant Number JPMXP1020230406).
We also acknowledge use of the software \textsc{Matplotlib}~\cite{Matplotlib2007}, \textsc{Numpy}~\cite{NumPy2020}, \textsc{SciPy}~\cite{SciPy2020}, and \textsc{IPython}~\cite{IPython2007}.

\bibliographystyle{JHEP}
\bibliography{Lib}
\onecolumngrid
\clearpage
\appendix

\setcounter{equation}{0}
\setcounter{figure}{0}
\setcounter{table}{0}
\setcounter{page}{1}
\makeatletter
\renewcommand{\theequation}{S\arabic{equation}}
\renewcommand{\thefigure}{S\arabic{figure}}
\renewcommand{\thepage}{S\arabic{page}}
\renewcommand{\thetable}{S\arabic{table}}

\begin{center}
\textbf{\large Supplemental Material for the Letter}\\
\textbf{\large Gravitational-Wave Signatures of Nonstandard Neutrino Properties\\ in Collapsing Stellar Cores}
\end{center}

In this Supplemental Material, we provide additional information on (A)~resolution tests for our core-collapse simulations with FFC treatment, (B)~the GW analysis and corresponding results, and (C)~possible BSM particle scenarios with neutrino FC, whose basic effects we expect to be represented by our generic models employing a parametric implementation of FFC.

\section{A.~Higher-resolution simulations}

\begin{figure}[h!]
\centering
    \includegraphics[width=17cm]{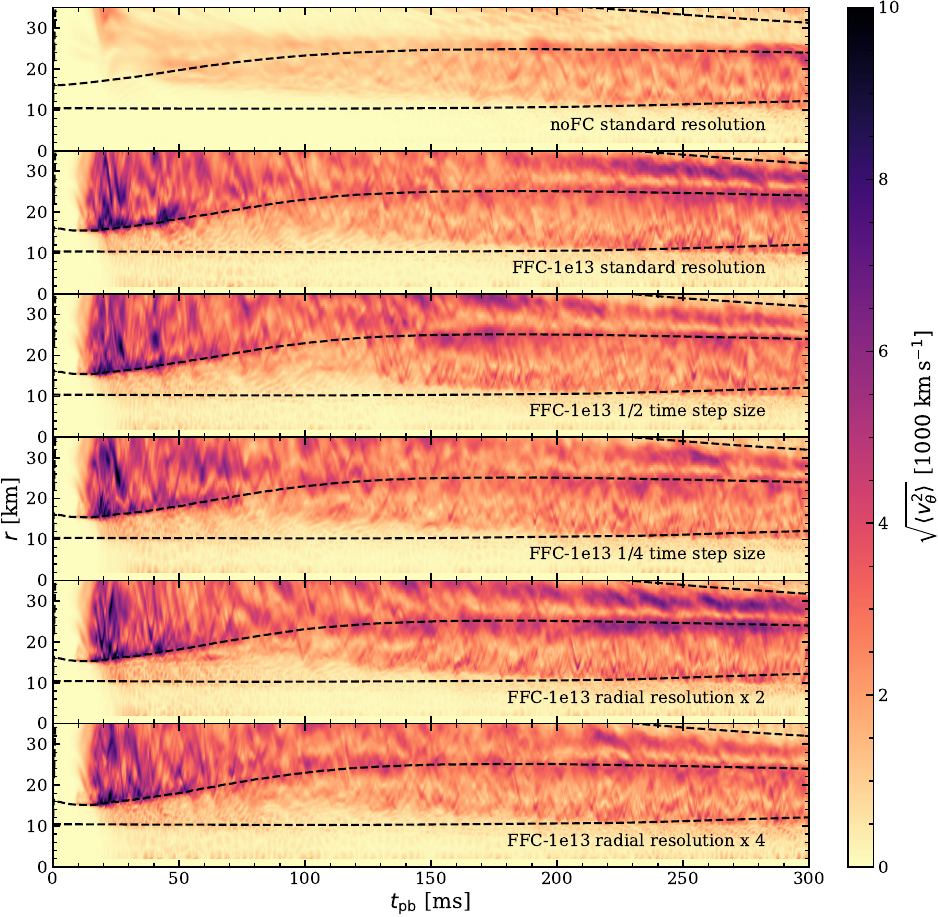}
    \caption{
    Time-step and radial resolution dependence of the evolution of PNS convection in our 20\,M$_\odot$ models with FC at $\rho < 10^{13}$\,\gccm.
    The different panels display the color-coded root of the (mass weighted) angular average of the squared lateral velocity (i.e., the standard deviation of the lateral velocity) as a function of post-bounce time in a radial domain up to 35\,km.
    The top panel shows the noFC case with our standard resolution and the second panel the run with FFC and standard resolution for comparison.
    The third and following panels from the top show our results with half the time step size, a quarter of the time step size, twice the number of radial grid zones, and four times the number of radial zones.
    The dashed lines mark, from deeper inside outward, the iso-density radii for $10^{14}$, $10^{13}$, and $10^{12}$\,\gccm\ for orientation.
    Increasing the time or radial resolution does not have any significant impact on the time when the PNS convection starts nor on the amplitude of the non-radial velocities beyond the expected stochastic variations and secondary resolution-dependent fine-scale flow structures.
    }
    \label{fig:app6}
\end{figure}

In our numerical neutrino-hydrodynamics simulations we treat neutrino FFC via an operator-splitting approach, assuming instantaneous neutrino equilibration (by pairwise $\nu\bar\nu$ conversion under the constraint of individual lepton-number conservation) in each time step within the spatial cells of our computational grid  after having performed the updates due to hydrodynamic advection and neutrino propagation and source terms.
When using an explicit, second-order Runge-Kutta time integrator with typical time-step size on the order of a few 100\,ns, we apply the flavor-mixing procedure also in each Runge-Kutta sub-step.
This implies that FC is also taken into account during the updating for gas advection and neutrino interactions.
Such an approach requires the validity of a scale separation between the tiny scales of FFC and the much coarser scales used in the space and time discretization of neutrino-hydrodynamic models.
A recent study in Ref.~\cite{Xiong:2024pue} has provided support for this ``coarse-grained'' approach in space and time for the typical spatial and temporal resolutions that are feasible in long-time global hydrodynamical simulations of explosive astrophysical events.
This study demonstrated that such an effective scheme yields results that quantitatively agree with those obtained by direct quantum kinetic simulations, and also collisional feedback effects could be precisely captured when FFC happens interior to the neutrinospheres, without the effective scheme resolving the small time and length scales of FFC.    

Nevertheless we ran test simulations initialized with the 20\,M$_\odot$ progenitor and for a FFC threshold density of $\rho_\mathrm{c} = 10^{13}$\,g\,cm$^{-3}$, using increased time resolution (by a factor 2 and 4) or increased radial resolution (also by a factor 2 and 4) in the region where FC makes an impact.
We performed this resolution study in 2D in order to test not only the consequences for FC but also the effects on PNS convection and the GW emission.
The smaller time stepping was realized by setting the Courant-Friedrichs-Lewy (CFL) factor~\cite{Courant1928a} to half or one quarter, respectively, of the value used in our reference simulations discussed in the main text.
For the radial resolution we increased the number of radial zones to twice or four times the grid cells applied in the reference simulations.
Since the most relevant effects happen in the convectively active region, we only refined the radial grid in this domain.
When constructing the finer grid we made sure that the size of neighboring zones did not change by more than 2.5\%.

Figure~\ref{fig:app6} displays color-coded the standard deviation of the lateral velocity ($\sqrt{\langle \mathrm{v}^2_{\theta} \rangle}$, with a mass-weighted angle average denoted by $\langle...\rangle$), thus visualizing the convective activity in the PNS between the grid center and 35\,km as a function of time.
Our noFC and FC-1e13 cases with standard resolution are compared with the different cases where we reduced the time-step size or increased the number of radial grid cells in the PNS convection layer.
In the first 150\,ms after bounce hardly any differences are visible beyond those expected by stochastic variations.
Later in the evolution one can witness enhanced peak velocities in certain time intervals, but not growing systematically with higher resolution.
These secondary differences are again explained by stochastic variations and might also originate from better resolved fine-scale flow structures.
In the crucial period for our discussion between core bounce and $\sim$150\,ms later, the onset of PNS convection, the extent of the convective region, and the strength of the convective activity measured by $\sqrt{\langle \mathrm{v}^2_{\theta} \rangle}$ do not reveal any significant differences between the FFC simulations with varied time and radial resolutions, in contrast to the clear differences compared to the noFC model.
Correspondingly, the GW signals agree very well between all cases with different resolutions, in particular regarding the dominant contributions to the GW frequency spectrum.

\begin{figure}[b]
\centering
    \includegraphics[width=1.\textwidth]{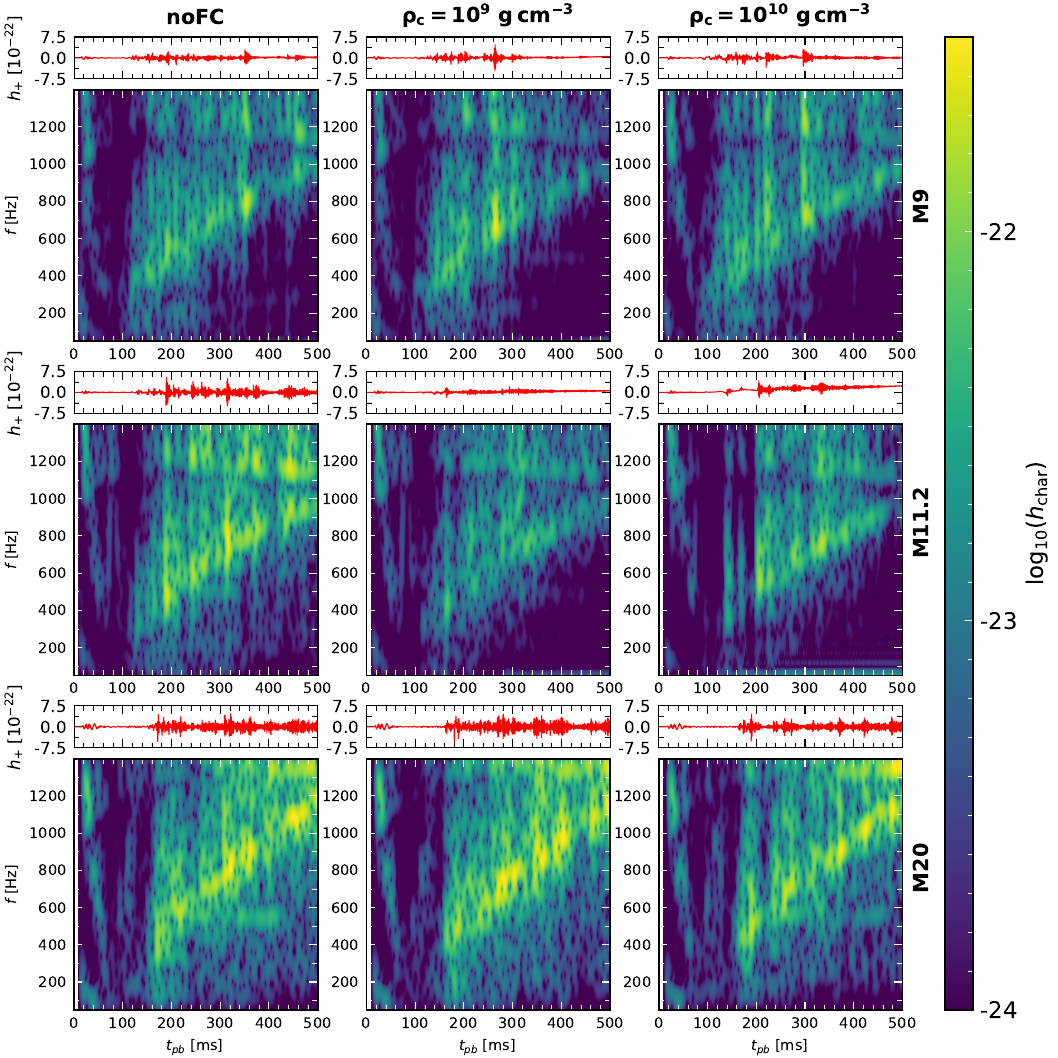}
    \caption{
    Analogue to Fig. 3.
    GW strains and their spectrograms for a SN distance of 10\,kpc as functions of time after bounce, shown for an observer at the equator.
    The different panels display these results for our 2D SN simulations of a 9\,M$_\odot$ progenitor ({\em top row}), 11.2\,M$_\odot$ star ({\em middle row}), and a 20\,M$_\odot$ case ({\em bottom row}), considering different FC scenarios: noFC ({\em left panels}) and FC at densities lower than $\rho_\mathrm{c}= 10^{9}$\,g\,cm$^{-3}$ ({\em panels in middle column}), and at $\rho < \rho_\mathrm{c}= 10^{10}$\,g\,cm$^{-3}$ ({\em right panels}).
    }
    \label{fig:GW-2}
\end{figure}

\begin{figure}
    \includegraphics[width=1.\textwidth]{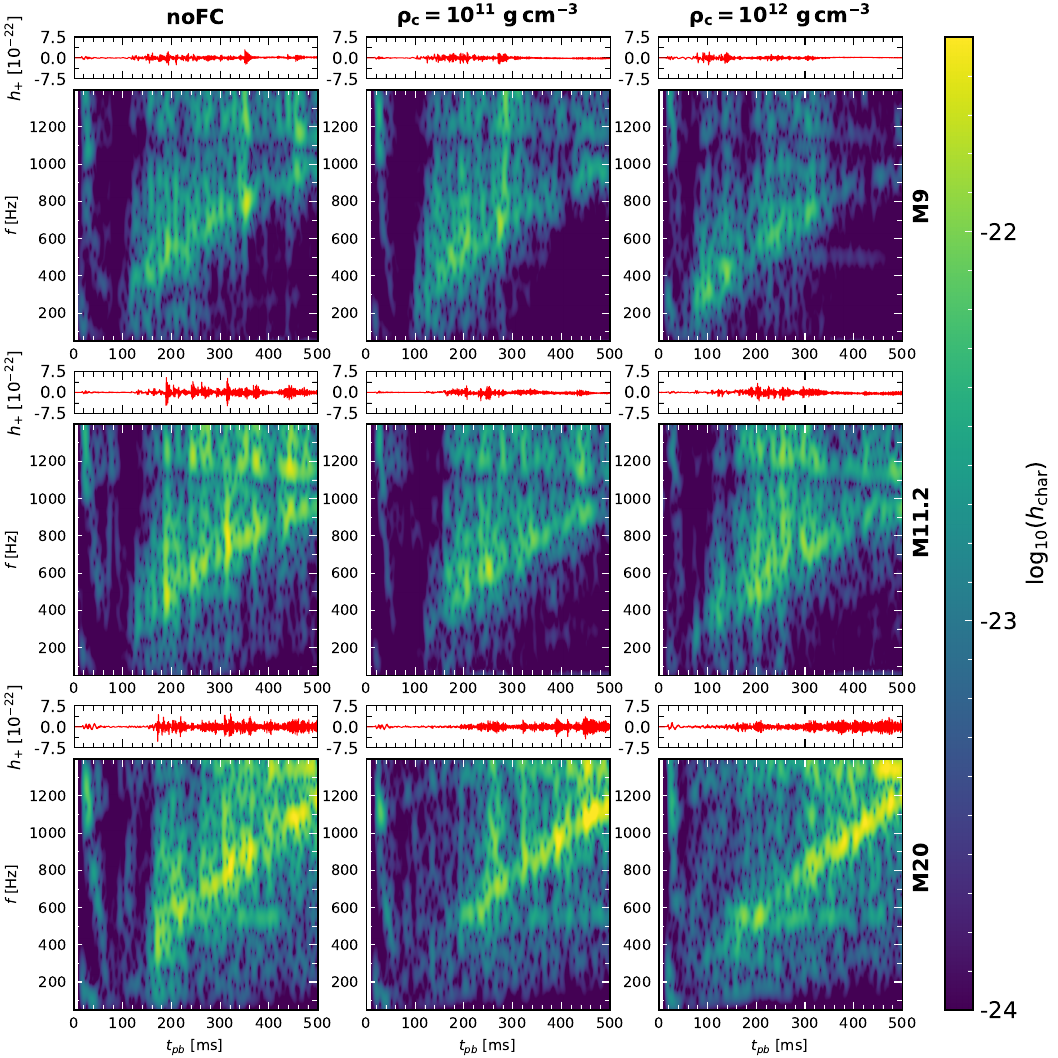}
    \caption{
    Analogue to Fig.~\ref{fig:GW-2}, but for SN simulations with FCs at $\rho < \rho_\mathrm{c}= 10^{11}$\,g\,cm$^{-3}$ and $\rho < \rho_\mathrm{c}= 10^{12}$\,g\,cm$^{-3}$ in the middle and right panels, respectively.
    }
    \label{fig:GW-3}
\end{figure}

\section{B.~Calculation of GW strains, spectrograms, and detectability}

\subsection{I.~GW signal calculation}

We follow Ref.~\cite{Finn1990a} to calculate the GW strain in the equatorial plane,
\begin{equation}
    h^\mathrm{equ}_+(f,\tau) = \frac{G}{c^4D} {\partial_t} \int dV \rho \bigl[v_r r (3\cos^2\theta-1) - 3 v_{\theta} r \cos\theta \sin\theta\bigr]\,,
\end{equation}
with $G$ being the gravitational constant, $c$ the speed of light, and $D$ the distance between source and observer.
$\rho$, $v_r$, $v_{\theta}$, $r$, and $\theta$ are the mass density, radial and polar velocity components, and the 2D spatial coordinates, respectively.
The derivation of this formula starts with the standard quadrupole formula and makes use of the continuity equation to eliminate one time derivative.

We calculate the spectrograms as laid out in Refs.~\cite{Kuroda+2014, Murphy+2009, Shibagaki2021a, Andresen2017a}.
The characteristic strain $h_{\mathrm{char}}$ is defined as
\begin{equation}
    h_{\mathrm{char}}=\sqrt{\frac{2}{\pi^2}\frac{G}{c^3}\frac{1}{D^2}\frac{\mathrm{d}E}{\mathrm{d}f}}
    \label{eq:hc}
\end{equation}
with $f$ being the frequency.
The spectral energy density d$E$/d$f$ is calculated using a Fourier transformation $\mathrm{FT\{\cdot\}}$.
\begin{equation}
    \frac{\mathrm{d}E}{\mathrm{d}f} = \frac{\pi}{4}\frac{c^3}{G}{D^2}{f^2}|\mathrm{FT}\{h^\mathrm{equ}_+\}|^2.
    \label{eq:Ef}
\end{equation}
For the spectrograms the time-dependent spectral energy density is obtained by a short-term Fourier transformation $\mathrm{STFT\{\cdot\}}$,
\begin{equation}
    \mathrm{STFT}\{A\}(f,\tau) = \int^{\infty}_{-\infty} A(t) H(t-\tau)e^{-2\pi i ft} \mathrm{d}t\,.
\end{equation}
The time dependence is achieved through the usage of a window function $H$ that is non-zero only in a narrow time interval.
We use a Hann window with width $\delta t$ = 20\,ms,
\begin{equation}
    H(t-\tau) = 
    \begin{cases}
        \frac{1}{2} \left[1+\mathrm{cos}(\frac{t-\tau}{\delta t}2\pi)\right] &\text{for } |t-\tau|\leq\frac{\delta t}{2}\,, \\
        0                                                                    &\text{for } |t-\tau|>   \frac{\delta t}{2}\,, \\
    \end{cases}
    \label{eq:Hann}
\end{equation}
with $t$ and $\tau$ being the time and offset time, respectively.

\begin{table}[b]
\caption{
SNR ratios based on the integral of Eq.~(\ref{eq:SNR}), evaluated with the characteristic strain for the first 100\,ms after bounce in our 2D models and with $f_\mathrm{min}=10$\,Hz and $f_\mathrm{max} =2000$\,Hz, assuming a source distance of 10\,kpc.
Following Eq.~(33) of Ref.~\cite{Andresen2017a}, a $\mathrm{SNR} \gtrsim 12$--16 is required for a detection when $f_\mathrm{max}$ is varied between 600\,Hz and 2000\,Hz.
Note that the signal above a frequency of $\sim$\,600\,Hz accounts for a minor contribution to the GW power spectrum (see Fig.~\ref{fig:GWspectra}) and the SNR.
}
\begin{ruledtabular}
\begin{tabular}{l c c c c c}
Model & advanced Virgo & advanced LIGO & KAGRA & Einstein Telescope & Cosmic Explorer \\
\hline
M9-noFC            &   0.7 &   1.0 &   1.0 &  14.7 &  18.3 \\
M9-FFC-1e13        &  10.3 &  14.3 &  13.0 & 199.4 & 248.9 \\
M11.2-noFC         &   0.6 &   0.8 &   0.7 &  10.7 &  14.0 \\
M11.2-FFC-1e13     &   8.7 &  12.0 &  11.1 & 166.7 & 213.8 \\
M20-noFC           &   1.7 &   2.4 &   2.5 &  35.5 &  44.4 \\
M20-FFC-1e13       &  10.4 &  14.2 &  11.9 & 186.2 & 241.6 \\
\end{tabular}
\end{ruledtabular}
\label{tab:SNR}
\end{table}

\begin{figure}[b]
    \includegraphics{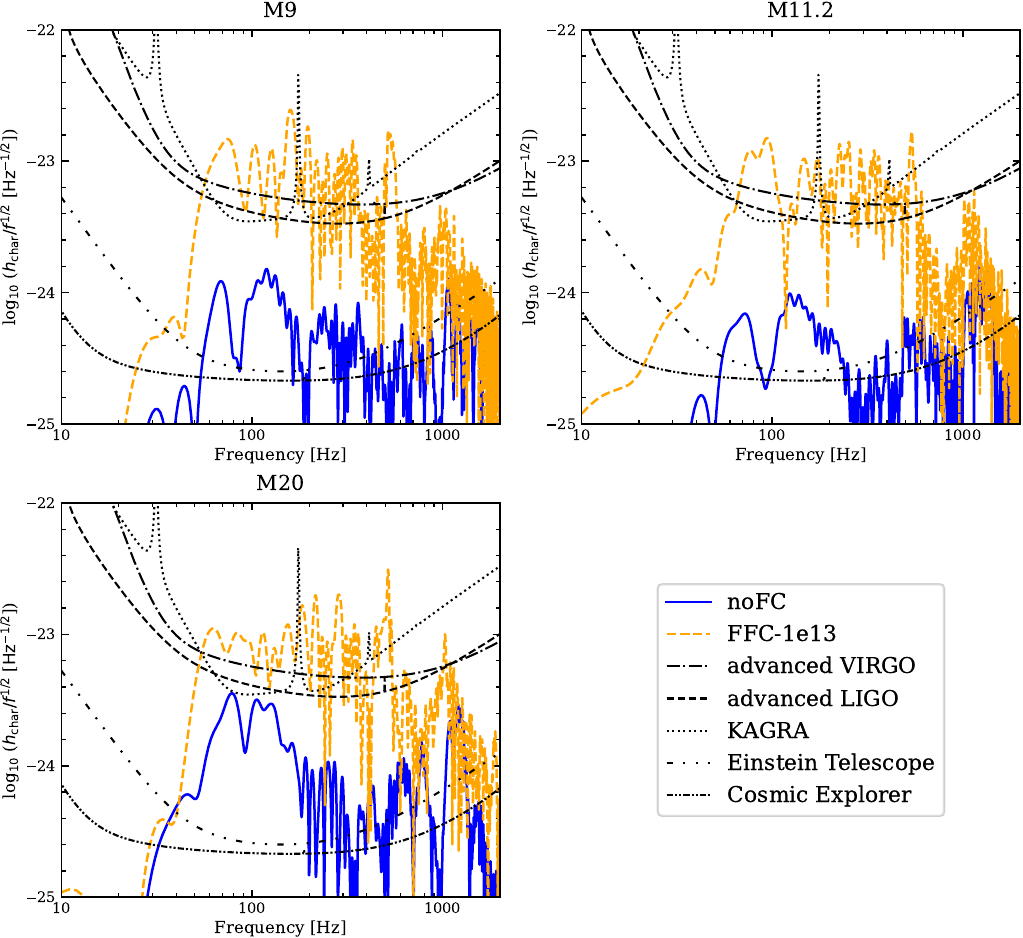}
    \caption{
    Comparison of the GW amplitude spectral densities and the amplitude spectra of the detector noise for different existing and future interferometer experiments (broken black lines).
    The signals in the FFC-1e13 simulations of the 9\,M$_\odot$, 11.2\,M$_\odot$, and 20\,M$_\odot$ models with enhanced GW emission (orange) are about one order of magnitude stronger than the signals in the noFC simulations (blue).
    }
    \label{fig:GWspectra}
\end{figure}

\subsection{II.~GW strains and spectrograms for additional neutrino FC models}

In Figures~\ref{fig:GW-2} and~\ref{fig:GW-3} we present the GW strains and their spectrograms for the models with $\rho_\mathrm{c}= 10^{9}$ and $10^{10}$\,g\,cm$^{-3}$ (Fig.~\ref{fig:GW-2}) and $10^{11}$ and $10^{12}$\,g\,cm$^{-3}$ (Fig.~\ref{fig:GW-3}).
In these simulations we consider FCs motivated by \textit{fast} FC as they can happen due to coherent forward scattering of neutrinos on other neutrinos.
The FCs occur outside the PNS or close to its surface and thus have no or only a mild influence on PNS convection.
For the sake of comparison, we also show the noFC model.
Note that all of these models exhibit extended periods, lasting 50--100\,ms, of relative quiescence, following a short, low-amplitude GW activity associated with prompt postshock convection right after core bounce (see also Fig. 3 and the corresponding discussion in the main text).
Hence only FCs in the deeper regions of the PNS in the cases with $\rho_\mathrm{c}= 10^{13}$ and $10^{14}$\,g\,cm$^{-3}$ lead to the significant GW emission in the window of relative quiescence.
These otherwise calmer periods end when hydrodynamic instabilities in the neutrino-heated postshock layer set in and the associated vigorous nonradial mass motions produce GW emission.
The duration of the more silent window is shortened in some models where FCs affect the neutrino emission properties such that stronger neutrino heating occurs and thus the hydrodynamic instabilities in the postshock volume develop more quickly.
A subset of models including FCs also develops faster explosions than without FCs, which weakens their GW emission after the onset of the explosion relative to the noFC cases.

\subsection{III.~GW signal detectability}

For our assessment of the detectability of the GWs we compare the amplitude spectral density of the signal with the noise of current (advanced VIRGO~\cite{Acernese2015a} and advanced LIGO~\cite{LSC2015a,Abbott2020c} and KAGRA~\cite{Aso2013a,Michimura2020a}) and next-generation GW detectors (Einstein Telescope~\cite{Hild2008_1} and Cosmic Explorer~\cite{Srivastava2022a}) (see Fig.~\ref{fig:GWspectra}).
To obtain the time-independent spectral energy density for this purpose, we do not apply the window function.
We evaluate the integral in the Fourier transform for $0\,\mathrm{ms} \le t \le 100\,\mathrm{ms}$, i.e., in the phase of relative quiescence in noFC models.

Following~\cite{Flanagan1998a} the signal-to-noise ratio (SNR) of a roughly isotropic source in an optimally oriented detector is given by
\begin{equation}\label{eq:SNR}
    (\mathrm{SNR})^2 = \int_{f_{\mathrm{min}}}^{f_{\mathrm{max}}} df
    \frac{h_{\mathrm{char}}^2}{f^2 S(f)}\,.
\end{equation}
Here $f_{\mathrm{min}}$ and $f_{\mathrm{max}}$ are the boundaries of the frequency interval used for the analysis, and $S(f)$ is the power-spectral density of the detector noise.
For a detection the SNR needs to be roughly $\gtrsim 4.3 (\delta f \Delta t)^{1/4}$ (Eq.~(33) in~\cite{Andresen2017a}).
The SNRs of the GW signal for the models considered in this paper can be found in Table~\ref{tab:SNR}.
Note that simulations in axial symmetry (2D) overestimate the amplitude of the emitted GW signal roughly by a factor of 10, but overall the frequency range of the GW emission is not sensitive to the dimension of the numerical models~\cite{Andresen2017a, Mezzacappa2020a}.
Accordingly, we therefore expect the SNRs in 3D models to be similarly lower by about a factor of 10.
However, this has no impact on the increase of the SNR when comparing a model with FC inside the PNS (FFC-1e13) to the corresponding model with no flavor conversion (noFC).

\bigskip
\section{C.~Nonstandard scenarios of neutrino flavor conversion}

BSM neutrino properties offer several mechanisms for inducing FC inside the PNS with a potential impact on the PNS cooling evolution, PNS convection, and the corresponding GW emission.
In the following, we describe two possibilities where we expect consequences similar to those discussed on grounds of our generic numerical FC models presented in the main text, namely nonstandard neutrino self-interaction ($\nu$NSSI) and resonant conversion of sterile neutrinos inside the PNS.

\subsection{I.~Flavor conversion by nonstandard neutrino-neutrino interaction}

Our first example is $\nu$NSSI~\cite{Bialynicka-Birula:1964ddi, Bardin:1970wq, Berryman:2022hds} that can drive flavor equipartition between neutrinos inside the PNS~\cite{Abbar:2022jdm}.
In particular, a vector mediator described by the effective Lagrangian $\mathcal{L}_{\mathrm{eff}} \supset G_{\mathrm{F}} [\mathsf{G}^{\alpha\beta} \bar{\nu}_\alpha \gamma^{\mu} \nu_\beta] [\mathsf{G}^{\xi\eta} \bar{\nu}_\xi \gamma_{\mu} \nu_\eta]$ results in new interaction terms that couple neutrinos of different flavors through $\mathsf{G}^{\alpha\beta}$~\cite{Abbar:2022jdm, Blennow:2008er}.
In the context of SN neutrino FCs, these new interaction terms are particularly significant because they permit off-diagonal couplings between neutrino flavors, introducing the potential for strong instabilities.
It has been demonstrated that in the presence of $\nu$NSSI, FCs can occur on scales~$\sim \xi^{-1}$ with $ \xi = \sqrt{2} G_{\mathrm{F}} |\mathsf{G}^{\alpha\beta}| n_\nu $ for certain values of $\alpha$ and $\beta$, provided that $\xi \gg \omega$, with $\omega$ being the typical neutrino vacuum oscillation frequency~\cite{Abbar:2022jdm}.

\begin{figure}
\centering
    \includegraphics{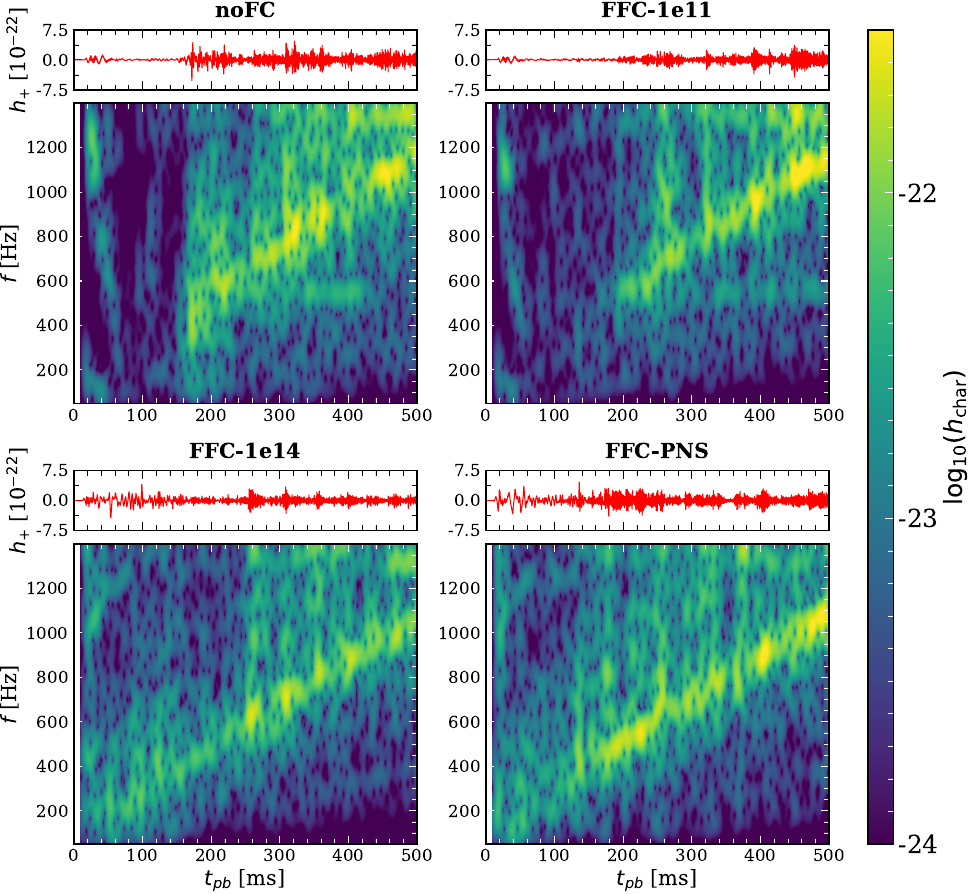}
    \caption{
    Analogue to Fig.~\ref{fig:GW-2} and Fig. 3, but only for SN simulations with the 20\,M$_\odot$ progenitor and including a simulation with FC only applied inside the PNS, i.e., when $\rho>\rho_{\mathrm{c}}=10^{11}\gccm$ (FFC-PNS).
    FC is still limited to be lepton number preserving.
    Note that the results for FFC-PNS are very similar to FFC-1e14 in the first $\sim$150\,ms after bounce.
    At later times the GW strain resembles more closely the noFC case, because FC does not occur exterior to the PNS.
    }
    \label{fig:FCinPNS}
\end{figure}

Current constraints on $\nu$NSSI are relatively weak and highly model-dependent.
Specifically, $|\mathsf{G}^{\alpha\beta}|$ could be several orders of magnitude larger than unity~\cite{Berryman:2022hds}.
Naturally, to be consistent with the neutrino treatment in our generic FC modeling guided by the case of FFC, we can only consider BSM scenarios here in which the impact on the scattering cross sections is negligible.
This means that our study does not yield self-consistent constraints for cases where the nonstandard part of the neutrino flavor-coupling terms fulfills $|\mathsf{G}^{\alpha\beta}| \gtrsim \mathcal{O}(1)$, because such strong BSM weak interactions must also be taken into account in the neutrino interaction rates.
Of course, this stronger BSM physics is another case for which our pioneering study motivates further investigations.

One of the interesting aspects of FCs induced by $\nu$NSSI is that their occurrence is primarily determined by the neutrino number density.
This fact makes such FCs largely consistent with the approach adopted in our study, which relies on a matter-density-based criterion for the presence of FC.
The alignment of both is justified by the close one-to-one correspondence between matter density and neutrino density up to the neutrinospheric region of the PNS, where neutrinos decouple from the stellar background.
Another noteworthy feature of $\nu$NSSI-induced FCs is that they can occur on relatively short time scales, much shorter than typical scales of neutrino scattering in the PNS interior.
As demonstrated in Ref.~\cite{Xiong:2024pue}, when FCs take place on sufficiently short (time and length) scales, the neutrino quantum kinetic effects can, in principle, be treated effectively by simple analytic prescriptions in a coarse-grained manner in global simulations that do not resolve the short scales of the FCs.
This means that the scale separation of the problem allows one to implement the asymptotic or equilibrium state of the neutrino gas established by FCs via an effective scheme coupled to the traditional neutrino transport treatment including collisions and advection, without solving the full quantum kinetic transport equations.
As mentioned already in Section~A of the SM, the results of Ref.~\cite{Xiong:2024pue} strongly support the methodology employed in our study.

In order to adapt our generic neutrino FC treatment even closer to the $\nu$NSSI situation, we present in Fig.~\ref{fig:FCinPNS} the results of a 20\,M$_\odot$ simulation in which we assumed that FCs happen throughout the entire PNS, i.e., in regions where $\rho \geq 10^{11}$\,g\,cm$^{-3}$.
In contrast to the other simulations discussed in this work, where $\rho < \rho_\mathrm{c}$ holds for the FC domain, the threshold density here serves as a lower density limit of the region where FCs take place.
Below the density $\rho_\mathrm{c}$, FCs are assumed not to occur any longer, consistent with the expectations for $\nu$NSSI.
The good agreement of the GW signal for the first $\sim$150\,ms after core bounce between this case (FFC-PNS) and the previous FFC-1e14 model (Fig.~\ref{fig:FCinPNS}) provides further evidence that the key phenomenon we describe does not critically depend on the specific implementation details of the FC condition.
In typical core-collapse SN models, the neutrino number density near the PNS surface (close to the neutrinospheres) is on the order of $n_{\nu} \gtrsim 10^{32}$\,cm$^{-3}$ during the accretion phase.
This translates into $\xi/\omega \gtrsim 10^5 |\mathsf{G}^{\alpha\beta}|$ as a lower limit for the entire PNS interior, because $n_\nu$ increases deeper inside the PNS.
This is consistent with the condition $\xi/\omega\gg 1$ stated in the first paragraph, if $|\mathsf{G}^{\alpha\beta}| \gtrsim 10^{-4}$.

Assuming that a FC-induced GW burst is not observed, very stringent constraints on nonstandard neutrino properties in the context of $\nu$NSSI could be placed.
Specifically, one could exclude $\mathcal{O}(10^{-6}) \lesssim |\mathsf{G}^{\alpha\beta}| \lesssim \mathcal{O}(1)$ for certain values of $\alpha$ and $\beta$~\cite{Abbar:2022jdm}.
Here we have assumed that during the GW-calm phase in the noFC model, the neutrino number densities are $n_\nu \gtrsim 10^{34}$\,cm$^{-3}$ for $\rho \gtrsim 10^{12}$\,g\,cm$^{-3}$.
We repeat that our study does not yield self-consistent constraints for $|\mathsf{G}^{\alpha\beta}| \gtrsim \mathcal{O}(1)$, since strong BSM weak interactions must be incorporated into the neutrino interaction rates in this range of $|\mathsf{G}^{\alpha\beta}|$.
In addition, for very large values of $|\mathsf{G}^{\alpha\beta}|$, one should consider the intriguing possibility of a neutrino fluid in the SN environment~\cite{Chang:2022aas, Fiorillo2024a, Fiorillo2024c, Bhattacharya:2023wzl}.

\begin{figure}[b]
\centering
    \includegraphics{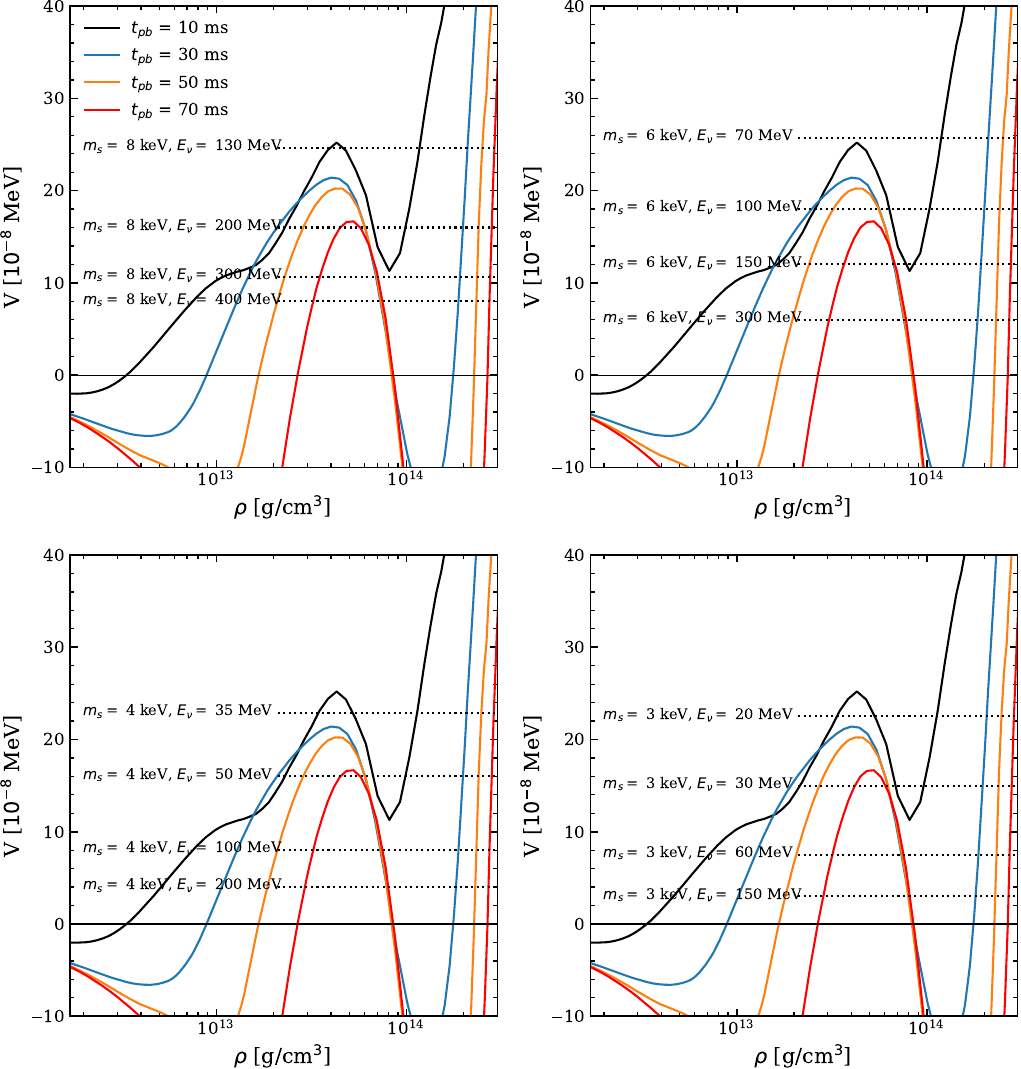}
    \caption{
    Forward-scattering potential of $\nu_{e}$ (as defined in Eq.~\ref{eq:V}) as a function of density for the early post-bounce phase in model M20-noFC (colored lines, same for each panel).
    MSW resonances with sterile neutrinos can occur, if the potential is positive.
    Intersection points with the dotted lines mark the radii where the resonance condition is met for neutrinos of the indicated energies.
    The different panels show resonances for different sterile neutrino masses of 8, 6, 4, and 3\,keV (from top left to bottom right).
    For small mixing, the resonance condition reads $V=m^2_s/(2E_\nu$), that is, e.g., sterile neutrinos of mass 6\,keV are in resonance with active neutrinos of energy 100\,MeV, if $V=18\times10^{-8}$\,MeV.
    }
    \label{fig:sterilenu}
\end{figure}

\subsection{II.~Neutrino FC by MSW resonances with sterile neutrinos}

Apart from $\nu$NSSI, which can induce FC inside the PNS and thereby can be expected to boost convective activity in the PNS interior, other scenarios can also produce similar effects.
One notable example is the resonant MSW active-sterile neutrino conversion ($\nu_e \rightleftharpoons \nu_s$) within the PNS, as discussed in Refs.~\cite{Hidaka:2007se, Hidaka:2006sg, Caldwell:1999zk, Warren:2014qza, Warren:2016slz, Syvolap:2019dat,Suliga:2020vpz, Ray:2023gtu,Ray:2024jeu}.
In particular, we consider sterile neutrinos with keV masses (relevant as viable dark-matter candidates) and neutrinos with particle energies $E_\nu$ between some 10\,MeV up to a few 100\,MeV.
This scenario involves:
(1)~Coherent  $\nu_e \rightarrow \nu_s$ conversion deep in the core at the first MSW resonance, where $\nu_e$ energies are high ($E_\nu\sim$100\,MeV) at densities around (1--$3)\times 10^{14}$\,g\,cm$^{-3}$;
(2)~outward streaming of high-energy $\nu_s$ nearly at the speed of light; and
(3)~energy deposition via $\nu_s \rightarrow \nu_e$ reconversion and subsequent absorption of the high-energy $\nu_e$ at a second MSW resonance well interior to the $\nu_e$-neutrinosphere at densities between several $10^{13}$\,g\,cm$^{-3}$ and $<$\,$10^{14}$\,g\,cm$^{-3}$.

The resonant $\nu_e \rightleftharpoons \nu_s$ conversion occurs when 
\begin{equation}
    V = \cos2\theta_{\nu_s} \frac{\Delta m^2}{2E_\nu}\,,
\label{eq:resonance}
\end{equation}
where $\theta_{\nu_s}$ is the tiny effective vacuum mixing angle ($\cos2\theta_{\nu_s} \simeq 1$) and $\Delta m^2 = m_2^2 - m_1^2 \simeq m_s^2$.
In Eq.~(\ref{eq:resonance}) the $\nu_e$ forward-scattering potential is given by  
\begin{equation}\label{eq:V}
     V = \sqrt{2} G_{\mathrm{F}} (n_e-\textstyle{\frac{1}{2}}\,n_n) + 2 \sqrt{2} G_{\mathrm{F}} (n_{\nu_e}-n_{\bar\nu_e})\,,
\end{equation}
with $n_e$ and $n_n$ being the net electron number density and the total neutron number density.
Here, we have assumed $n_{\nu_x} = n_{\bar\nu_x}$, so that there is no net contribution to the potential $V$ from heavy-lepton neutrinos.

Although the detailed physics of this scenario is not exactly compatible with the FFC assumptions in our simulated models, the local energy deposition by enhanced $\nu_e$ absorption near the MSW resonance below $10^{14}$\,g\,cm$^{-3}$ (as~illustrated in Fig.~\ref{fig:sterilenu} for various neutrino masses and energies) provides another example where BSM physics is likely to induce PNS convection early after core bounce (similar to the $\rho_\mathrm{c} = 10^{14}\,$g\,cm$^{-3}$ case in Fig. 2).
Here, sterile neutrinos are involved, which enhance the energy transport from the PNS core to PNS layers farther out, where convectively unstable conditions can build up, thus leading to earlier and stronger convective activity in the interior of the PNS than expected in the absence of this FC effect.

While this scenario is not directly explored by our study, because it requires a different implementation of the neutrino FC physics in the SN simulations, our work provides strong motivation to revisit such a sterile-neutrino scenario to assess its impact on the GW signal from the next galactic stellar core-collapse event.

\end{document}